\documentclass[aps,prd,11pt,twoside,tightenlines,nofootinbib,showpacs,preprint,superscriptaddress]{revtex4-1}
\usepackage[colorlinks=true]{hyperref}
\usepackage{xcolor}
\hypersetup{colorlinks=true, citecolor=blue, urlcolor=blue, linkcolor=blue}
\usepackage{amsmath,amssymb}
\usepackage[final]{graphicx}
\usepackage{subcaption}
\usepackage{slashed}
\usepackage{float}
\usepackage[toc,page]{appendix}

\begin{document}

\title{Study of the resonance contributions in the $\Xi_{b}^{-} \rightarrow pK^{-}K^{-}$ decay}

\author{Zhong-Yu Wang}
\email{zhongyuwang@foxmail.com}
\affiliation{School of Physical Science and Technology, Lanzhou University, Lanzhou 730000, China}
\affiliation{Lanzhou Center for Theoretical Physics, Lanzhou University, Lanzhou 730000, China}

\author{Si-Qiang Luo}
\email{luosq15@lzu.edu.cn}
\affiliation{School of Physical Science and Technology, Lanzhou University, Lanzhou 730000, China}
\affiliation{Lanzhou Center for Theoretical Physics, Lanzhou University, Lanzhou 730000, China}

\author{Zhi-Feng Sun}
\email{sunzf@lzu.edu.cn}
\affiliation{School of Physical Science and Technology, Lanzhou University, Lanzhou 730000, China}
\affiliation{Lanzhou Center for Theoretical Physics, Lanzhou University, Lanzhou 730000, China}
\affiliation{Research Center for Hadron and CSR Physics, Lanzhou University and Institute of Modern Physics of CAS, Lanzhou 730000, China}
\affiliation{Key Laboratory of Theoretical Physics of Gansu Province, and Frontiers Science Center for Rare Isotopes, Lanzhou University, Lanzhou 730000, China}

\author{C. W. Xiao}
\email{xiaochw@csu.edu.cn}
\affiliation{School of Physics and Electronics, Hunan Key Laboratory of Nanophotonics and Devices, Central South University, Changsha 410083, China}

\author{Xiang Liu}
\email{xiangliu@lzu.edu.cn}
\affiliation{School of Physical Science and Technology, Lanzhou University, Lanzhou 730000, China}
\affiliation{Lanzhou Center for Theoretical Physics, Lanzhou University, Lanzhou 730000, China}
\affiliation{Research Center for Hadron and CSR Physics, Lanzhou University and Institute of Modern Physics of CAS, Lanzhou 730000, China}
\affiliation{Key Laboratory of Theoretical Physics of Gansu Province, and Frontiers Science Center for Rare Isotopes, Lanzhou University, Lanzhou 730000, China}

\date{\today}

\begin{abstract}
	
The decay process $\Xi_{b}^{-} \rightarrow pK^{-}K^{-}$ is studied with the final state interaction approach by considering the contributions from the $S$-wave meson-baryon interactions, and also the intermediate state $\Lambda(1520)$ in the $D$-wave. The low-lying resonances $\Lambda(1405)$ and $\Lambda(1670)$ have significant contributions, which are both dynamically generated from the $S$-wave final state interactions with isospin $I=0$. Furthermore, the $\Lambda(1520)$ state also has important contributions from $D$-wave. With these resonances contributions, the experimental data of the lower $pK^{-}$ invariant mass distributions are well described. We also discuss the contribution of another resonance in the $S$-wave with isospoin $I=1$, which cannot be ignored. Moreover, some of the branching fractions obtained for the corresponding decay channels are consistent with the experimental measurements.

\end{abstract}

\maketitle

\section{Introduction}

The weak decays of charmed and bottomed hadrons can not only be used to explore the $CP$-violation phenomena and new physics beyond the standard model, but also be suitable for exploring the nature of the intermediate resonances. In particular, the three-body decays of charmed and bottomed hadrons provide a good chance to study the properties of the intermediate resonances, and also a good opportunity to observe new resonances in the invariant mass spectra of the final states. 
It is well known that in 2015 the LHCb Collaboration reported two pentaquarklike resonances, i.e., $P_{c}(4380)^{+}$ and $P_{c}(4450)^{+}$ in the $J/\psi p$ invariant mass spectrum of the $\Lambda_{b}^{0}\rightarrow J/\psi K^{-}p$ decay \cite{LHCb:2015yax,Aaij:2015fea}, which was confirmed by a model-independent analysis of the data \cite{LHCb:2016ztz} and in the $\Lambda_b^0 \to J/\psi p \pi^-$ decay~\cite{LHCb:2016lve}. Furthermore, using the data of Run I and Run II, in 2019 the LHCb Collaboration updated their results for the $\Lambda_{b}^{0}\rightarrow J/\psi K^{-}p$ decay, where in fact three clear narrow structures, i.e., $P_{c}(4312)^{+}$, $P_{c}(4440)^{+}$ and $P_{c}(4457)^{+}$, were found \cite{LHCb:2019kea}.
Therefore, in recent years a large number of three-body charmed and bottomed baryons' decays, e.g., $\Lambda_{c}$ and $\Lambda_{b}$ decays, are caught much attentions experimentally \cite{Belle:2015wxn,LHCb:2016rja,BESIII:2016ozn,LHCb:2017jym,LHCb:2017xtf,BESIII:2018qyg,Belle:2020xku}, where more discussions about the theoretical and experimental progress on this issue can be referred to the recent reviews \cite{Chen:2016qju,Hosaka:2016pey,Esposito:2016noz,Guo:2017jvc,Olsen:2017bmm,Brambilla:2019esw} and references therein. 
In Ref. \cite{Belle:2018lws}, two $\Xi$ states, $\Xi(1620)^0$ and $\Xi(1690)^0$, were found in the $\Xi_c^+ \rightarrow \Xi^-\pi^+\pi^+$ decay. 
In the $\Lambda_{b}^0\pi^+\pi^-$ mass spectrum, two narrow resonances $\Lambda_{b}(6146)^0$ and $\Lambda_{b}(6152)^0$ were reported in Ref. \cite{LHCb:2019soc} and another wide one was found in Ref. \cite{LHCb:2020lzx}.
Moreover, the three-body decays of the $\Xi_{b}$ state are also caught much attentions.
In 2017, the $\Xi_{b}^{-}\to J/\psi \Lambda K^-$ decay was firstly observed by the LHCb Collaboration \cite{LHCb:2017fwd} with the suggestion of Ref. \cite{Chen:2015sxa} to look for the hidden-charm pentaquark states with open strangeness as predicted in Refs. \cite{Wu:2010jy,Wu:2010vk,Santopinto:2016pkp,Chen:2016ryt}. 
Indeed in 2021, the $P_{cs}(4459)^{0}$ state was observed in the $J/\psi \Lambda$ invariant mass distributions of the $\Xi_{b}^{-}\to J/\psi \Lambda K^-$ decay \cite{LHCb:2020jpq}. In the same year, the $\Omega_{b}^{-}\rightarrow\Xi_{c}^{+}K^{-}\pi^{-}$ decay was investigated by the LHCb Collaboration \cite{LHCb:2021ptx}, where four structures were observed in the $\Xi_{c}^{+}K^{-}$ invariant mass distributions, i.e., $\Omega_{c}(3000)^{0}$, $\Omega_{c}(3050)^{0}$, $\Omega_{c}(3065)^{0}$ and $\Omega_{c}(3090)^{0}$, which were consistent with the previous measurements of the LHCb Collaboration \cite{LHCb:2017uwr} and the Belle Collaboration \cite{Belle:2017ext}. 
Besides, the LHCb Collaboration had performed the amplitude analysis in Ref. \cite{LHCb:2021enr} for the decay $\Xi_{b}^{-}\rightarrow pK^{-}K^{-}$, which was reported for the first time in 2017 \cite{LHCb:2016hha}, and where the $CP$-violation effect was discussed and the contributions from the resonances $\Sigma(1385)$, $\Lambda(1405)$, $\Lambda(1520)$, etc., were found.
In the present work, this three-body decay, i.e., $\Xi_{b}^{-}\rightarrow pK^{-}K^{-}$, catches our interests to study the nature of the intermediate resonances in this decay process.

Indeed, the motivation of studying the $\Xi_{b}^{-}\rightarrow pK^{-}K^{-}$ decay \cite{LHCb:2016hha} was to look for $CP$-violation in the $b$-baryon decays, which was also concerned in theories \cite{Zhang:2021fdd,Sinha:2021mmx}. Even though the $CP$-violation effect was not found finally with the amplitude analysis in Ref. \cite{LHCb:2021enr}, as a by-product they reported that the resonance $\Sigma(1385)$, $\Lambda(1405)$, $\Lambda(1520)$, etc., had significant contributions to the decay amplitude, and especially the ones of  $\Lambda(1520)$ and  $\Lambda(1670)$ were observed with significance more than $5\sigma$, which motivate us to investigate theoretically the contributions of the states $\Lambda(1405)$ and $\Lambda(1670)$ in the present work.
Although the $\Lambda(1405)$ resonance was predicted and observed more than 60 years ago \cite{Dalitz:1959dn,Dalitz:1960du,Alston:1961zzd}, its structure and properties are still under debate. It was considered to be the normal three-quark baryon in the quark model \cite{Isgur:1978xj, Capstick:1986ter, Loring:2001ky}, but the mass obtained was higher than the experimental result \cite{Crede:2013kia}. On the other hand, it was confusing that the mass of the $\Lambda(1405)$ was significantly lighter than the lowest non-strange negative parity baryons $N(1520)$ \cite{Nakamura:2008zzc}. 
Meanwhile, the $\Lambda(1405)$ was dynamically produced in the coupled channels interactions \cite{Kaiser:1995eg, Kaiser:1996js,Oset:1997it} based on the interaction potentials from the chiral dynamics, where the experimental data for the cross sections were well described. 
Note that, Ref. \cite{Oset:1997it} only used one free parameter in the loop functions, where the on-shell approximations were taken \cite{Oller:1997ti}, and obtained consistent results for the cross sections and the other experimental data. 
Later, with the same method, which is also called as the chiral unitary approach (ChUA) \cite{Oller:2000ma, Oller:2000fj, Hyodo:2008xr, Oset:2008qh}, the $\Lambda(1670)$ state was also dynamically generated in the strangeness $S=-1$ and the isospin $I=0$ sector \cite{Oset:2001cn} and assumed as a bound state of the $K\Xi$ channel. 
Remarkably the two-pole structure for the $\Lambda(1405)$ state was found for the first time in the coupled channel interactions with the quark bag model \cite{Fink:1989uk}. Furthermore, the two-pole structure of the $\Lambda(1405)$ was investigated in details using the ChUA in Refs. \cite{Oller:2000fj,Jido:2002yz,Jido:2003cb,Garcia-Recio:2003ejq,Hyodo:2007jq,Ikeda:2011pi,Ikeda:2012au,Guo:2012vv,Wang:2021lth,Lu:2022hwm}. 
However, the cross-sections of the transition $K^{-}p\rightarrow\eta\Lambda$ were measured by the Crystal Ball Collaboration \cite{CrystalBall:2001uhc}, which supported the picture of three-quark baryon for the $\Lambda(1670)$ state. In the chiral quark model, in Ref. \cite{Zhong:2008km} the $\Lambda(1670)$ could also be treated as a three-quark state based on the analysis of the data of the $K^{-}p\rightarrow\pi^{0}\Sigma^{0}$ reaction \cite{Manweiler:2008zz}. Thus, the structures and properties of the $\Lambda(1405)$ and $\Lambda(1670)$ are still with a lot of controversies.

Recently, in Ref. \cite{Wang:2021lth} with the ChUA we systematically revisited the interactions of the $\bar KN$ and its coupled channels, and the single channel interactions of the channels $\bar KN$ and $\pi\Sigma$, respectively, where the nature of the $\Lambda(1405)$ was discussed that it was really two poles of the second Riemann sheet or two molecular states. The latest experimental results of the $\Xi_{b}^{-}\rightarrow pK^{-}K^{-}$ decay in Ref. \cite{LHCb:2021enr} can probe the properties of the resonances $\Lambda(1405)$, $\Lambda(1520)$ and $\Lambda(1670)$. Based on the two-body interaction results of Ref. \cite{Wang:2021lth}, we can investigate the $\Xi_{b}^{-}\rightarrow pK^{-}K^{-}$ decay with the final state interaction approach and try to hint the molecular nature of the states $\Lambda(1405)$ and $\Lambda(1670)$. 
More discussions about the molecular states can be found in the review of Ref. \cite{Guo:2017jvc}. 
Note that, since the $\Lambda(1520)$ state locates between the above two resonances and has significant contribution in the energy region that we are interested, we also take it into account in our formalism and ignore the less contribution resonances, such as  the ones $\Sigma(1385)$, $\Sigma(1775)$, $\Sigma(1915)$, and so on, as found in Ref. \cite{LHCb:2021enr}. 
Using a realistic chiral meson-baryon amplitude for the final state interactions (FSI), Ref. \cite{Miyahara:2018lud} predicted the line shapes of the $\pi\Sigma$ spectrum of the $\Xi_{b}^{0}\rightarrow D^{0}\pi\Sigma$ decay, which showed that the structure of the $\Lambda(1405)$ was destroyed by the interference between the direct generation and rescattering procedures. 
One can look forward to what we get from the $\Xi_{b}^{-}\rightarrow pK^{-}K^{-}$ decay with the FSI under the ChUA, which is a useful approach, such as a hidden charmed pentaquark state with strangeness predicted in the decay of $\Xi_{b}^{-}\to J/\psi \Lambda K^-$ in Ref. \cite{Chen:2015sxa} as mentioned above, which was confirmed by the experiment \cite{LHCb:2020jpq}. 
More discussions and applications for the heavy baryons three-body decays with the FSI based on the ChUA can be found in Refs. \cite{Miyahara:2015cja,Roca:2015tea,Xie:2016evi,Xie:2017gwc,Oset:2016lyh}. 

Our work is organized as follows. In Sec. \ref{sec:Formalism}, we will introduce the formulae of the decay amplitudes with the  FSI and the ChUA. Next, our results are shown in Sec. \ref{sec:Results}. At the end, a short conclusion is made in Sec. \ref{sec:Conclusions}.

\section{Formalism}
\label{sec:Formalism}

\begin{figure}[htbp]
	\centering
	\includegraphics[width=0.5\linewidth,trim=150 510 150 120,clip]{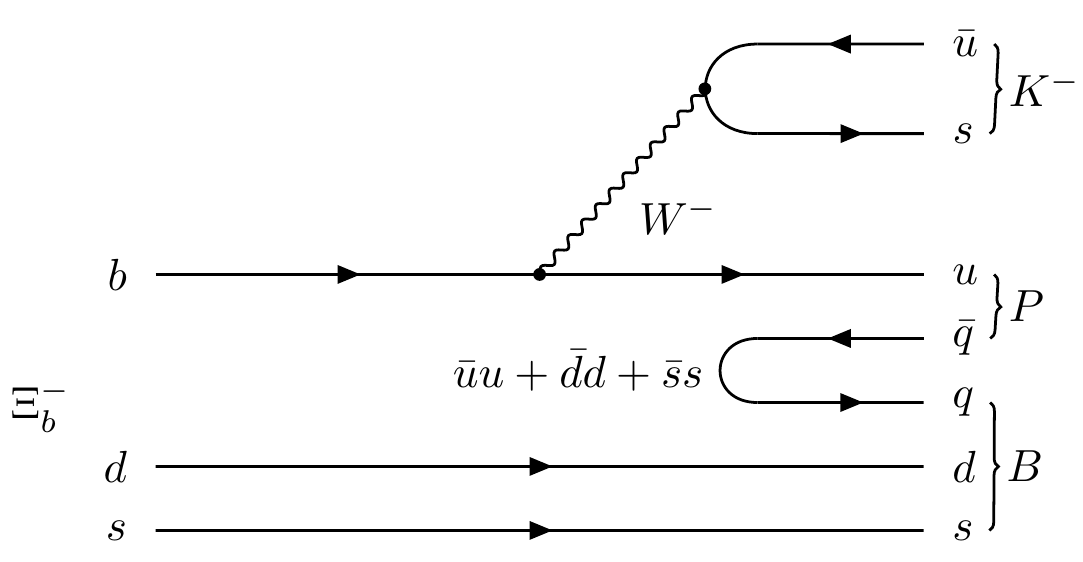} 
	\caption{The dominant diagram for the $\Xi_{b}^{-} \rightarrow pK^{-}K^{-}$ decay with the $W$-external emission mechanism, where $P$ and $B$ represent the pseudoscalar mesons and baryons, respectively.}
	\label{fig:Feynman1}
\end{figure} 

In the present work, we investigate the weak decay process of $\Xi_{b}^{-}\rightarrow pK^{-}K^{-}$, taking into account the FSI of $K^{-}p$ with its coupled channels. The most important contribution of the weak decay process comes from the $W$-external emission mechanism based on topological classification \cite{Chau:1982da, Chau:1987tk}. Thus, we only consider the dominant mechanism as shown in Fig. \ref{fig:Feynman1}, and omit the other contributions such as $W$-internal emission, $W$-exchange, $W$-annihilation, etc.
\footnote{In fact, these mechanisms can be obtained by rearranging the quark lines of Fig. \ref{fig:Feynman1} or considering the absorption diagrams \cite{Chau:1982da}, see more discussions in Refs. \cite{Miyahara:2015cja, Miyahara:2016yyh}.}, 
which are suppressed by the color factor \cite{Miyahara:2015cja, Roca:2020lyi}. As shown in Fig. \ref{fig:Feynman1}, the $ds$ quark pair in the $\Xi_{b}^{-}$ has spin $S = 0$ and has the flavor wave function $\frac{1}{\sqrt{2}}(ds-sd)$, which is the most attractive ``good" diquark and can be assumed as a spectator in the weak decay process \cite{Miyahara:2015cja, Pavao:2017cpt}. Meanwhile, the $b$ quark in the $\Xi_{b}^{-}$ decays into the $u$ quark via an external emission $W^{-}$ boson. Then the $W^{-}$ boson creates the $\bar{u}$ and $s$ quarks, which eventually form a $K^{-}$ meson. The remaining quarks $\frac{1}{\sqrt{2}}u(ds-sd)$ are hadronized by introducing the quark pairs $(\bar u u+\bar d d+\bar s s)$ from the vacuum to form a meson and a baryon, as depicted in Fig. \ref{fig:Feynman1}. We know that the $\Xi_{b}^{-}$ state has the quark components with a flavor function as
\begin{equation}
	\begin{aligned}
		|\Xi_{b}^{-}\rangle\equiv\frac{1}{\sqrt{2}}|b(ds-sd)\rangle, 
	\end{aligned}
	\label{eq:Xib}
\end{equation}
and after the $b$ quark decays into the $u$ quark, we have the $uds$ cluster, proceeded as 
\begin{equation}
	\begin{aligned}
		|H\rangle=V_{P}V_{ub}V_{us}\frac{1}{\sqrt{2}}|u(ds-sd)\rangle, 
	\end{aligned}
	\label{eq:H1}
\end{equation}
where $V_{P}$ represents the vertex factor of the weak decay for the $\bar{q}q$ pair creation, which is assumed to be a constant independent on the invariant mass in our calculation. More details can be seen in Refs. \cite{Liang:2014tia, Ahmed:2020qkv}. The $V_{q_{1}q_{2}}$ represents the element of the Cabibbo-Kobayashi-Maskawa (CKM) matrix for the transition of $q_{1}\rightarrow q_{2}$ quarks. With the formation of $K^{-}$ meson by the $\bar{u}$ and $s$ quarks from the $W^{-}$ boson, the spectators, the $d$ and $s$ quarks, become a part of the light baryon and the $u$ quark from $b$ decay becomes a part of the meson combining the $\bar{q}q$ pair generated from the vacuum. These processes are formulated as
\begin{equation}
	\begin{aligned}
		|H\rangle & = V_{P}V_{ub}V_{us}\frac{1}{\sqrt{2}}|u(\bar{u} u+\bar{d} d+\bar{s} s)(ds-sd)\rangle \\
		&=V_{P}V_{ub}V_{us}\frac{1}{\sqrt{2}} \sum_{i=1}^{3}\left|P_{1 i} q_{i}(ds-sd)\right\rangle,
	\end{aligned}
	\label{eq:H2}
\end{equation}
where the $q_{i}$ is the quark field and the $P_{ij}$ is the $q\bar{q}$ pair matrix element as follows
\begin{equation}
	\begin{aligned}
		q \equiv\left(\begin{array}{l} u \\ d \\ s \end{array}\right), \quad P=\left(\begin{array}{lll}{u \bar{u}} & {u \bar{d}} & {u \bar{s}} \\ {d \bar{u}} & {d \bar{d}} & {d \bar{s}} \\ {s \bar{u}} & {s \bar{d}} & {s \bar{s}}\end{array}\right).
	\end{aligned}
	\label{eq:qP}
\end{equation}

The $SU(3)$ matrices for the pseudoscalar mesons and the lowest-lying baryon octet can also be represented by the corresponding hadron fields, i.e.,

\begin{equation}
	\begin{aligned}
		P=\left(\begin{array}{ccc}{\frac{1}{\sqrt{2}} \pi^{0}+\frac{1}{\sqrt{3}} \eta+\frac{1}{\sqrt{6}} \eta^{'}} & {\pi^{+}} & {K^{+}} \\ {\pi^{-}} & {-\frac{1}{\sqrt{2}} \pi^{0}+\frac{1}{\sqrt{3}} \eta+\frac{1}{\sqrt{6}} \eta^{'}} & {K^{0}} \\ {K^{-}} & {\bar{K}^{0}} & {-\frac{1}{\sqrt{3}} \eta}+\sqrt{\frac{2}{3}} \eta^{'}\end{array}\right),
	\end{aligned}
	\label{eq:pseudoscalar}
\end{equation} 

\begin{equation}
	\begin{aligned}
		B=\left(\begin{array}{ccc}
			\frac{1}{\sqrt{2}} \Sigma^{0}+\frac{1}{\sqrt{6}} \Lambda & \Sigma^{+} & p \\
			\Sigma^{-} & -\frac{1}{\sqrt{2}} \Sigma^{0}+\frac{1}{\sqrt{6}} \Lambda & n \\
			\Xi^{-} & \Xi^{0} & -\frac{2}{\sqrt{6}} \Lambda
		\end{array}\right).
	\end{aligned}
	\label{eq:baryon}
\end{equation}

\begin{figure}
	\centering
	\includegraphics[width=0.5\linewidth,trim=150 540 180 130,clip]{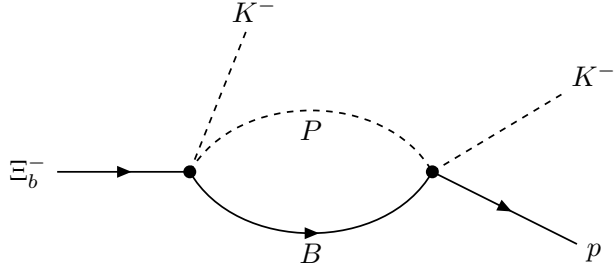} 
	\caption{The diagram for the pseudoscalar meson ($P$) and baryon ($B$) final state interactions via the rescattering mechanism.}
	\label{fig:Scatter1}
\end{figure}

\noindent
In Eq. (\ref{eq:pseudoscalar}), we take the standard mixing of the $\eta$ and $\eta^{'}$ in terms of a singlet and an octet of $SU(3)$. Then the hadronization processes in the quark level can be accomplished to the hadron level in terms of a pseudoscalar meson and a baryon, we obtain the final meson-baryon states for the hadronization procedure, given by

\begin{equation}
	\begin{aligned}
		|H\rangle = & V_{P}V_{ub}V_{us}\left(\frac{1}{2}|\pi^{0}\Sigma^{0}\rangle+\frac{1}{2\sqrt{3}}|\pi^{0}\Lambda\rangle
		+\frac{1}{\sqrt{6}}|\eta\Sigma^{0}\rangle
		\right.\\ & \left.+\frac{1}{3\sqrt{2}}|\eta\Lambda\rangle+|\pi^{+}\Sigma^{-}\rangle
		+|K^{+}\Xi^{-}\rangle\right),
	\end{aligned}
	\label{eq:H3}
\end{equation}

\noindent
where we neglect the irrelevant $\eta^{'}$ state since it is too massive and has no effect to the low energy region in the present work. The flavour functions of the mesons and baryons we used are the same as the ones in the appendix of Ref. \cite{Miyahara:2016yyh}. It is obvious that there is no $K^{-}p$ state directly produced in the tree level of the $\Xi_{b}^{-}$ decay based on the mechanism of Fig. \ref{fig:Feynman1}. However, the final meson-baryon states can go to further FSI, as depicted in Fig. \ref{fig:Scatter1}. The corresponding amplitudes with the contributions from the rescattering mechanism can be written as

\begin{equation}
	\begin{aligned} 
		\mathcal M_{S-wave}(M_{12},M_{13})=&\mathcal D \left[\frac{1}{2}G_{\pi^{0}\Sigma^{0}}(M_{12}) T_{\pi^{0}\Sigma^{0} \rightarrow K^{-}p}(M_{12})
		\right.\\ & \left.+\frac{1}{2\sqrt{3}}G_{\pi^{0}\Lambda}(M_{12}) T_{\pi^{0}\Lambda \rightarrow K^{-}p}(M_{12})
		\right.\\ & \left.+\frac{1}{\sqrt{6}}G_{\eta\Sigma^{0}}(M_{12}) T_{\eta\Sigma^{0} \rightarrow K^{-}p}(M_{12})
		\right.\\ & \left.+\frac{1}{3\sqrt{2}}G_{\eta\Lambda}(M_{12}) T_{\eta\Lambda \rightarrow K^{-}p}(M_{12})
		\right.\\ & \left.+G_{\pi^{+}\Sigma^{-}}(M_{12}) T_{\pi^{+}\Sigma^{-} \rightarrow K^{-}p}(M_{12})
		\right.\\ & \left.+G_{K^{+}\Xi^{-}}(M_{12}) T_{K^{+}\Xi^{-} \rightarrow K^{-}p}(M_{12})
		\right.\\ & \left.+(2\leftrightarrow3)\right],
	\end{aligned}
	\label{eq:amplitudes}
\end{equation}

\noindent
where $\mathcal D$ is a free parameter, which can be determined later by fitting the experimental data and have absorbed the vertex factor $V_{P}$, the elements of CKM matrix $V_{ub}$, $V_{us}$, and a global constant $C$ to match the events of the experimental data. Note that, we use the label $1$ for the proton, the label $2$ for the $K^{-}$ from the same vertex as the proton, and the label $3$ for the other $K^{-}$ directly created by the $W$-boson. The symbol $(2\leftrightarrow3)$ means exchanging the labels $2$ and $3$ in the amplitude $\mathcal{M}_{S-wave}$, which represents the symmetry of identical particles $K^{-}K^{-}$ in the final states. $M_{ij}$ is the energy of two particles in the center-of-mass frame. $G_{PB}$ and $T_{PB\rightarrow P^{'}B^{'}}$ are the loop functions and the scattering amplitudes, respectively, which will be introduced later and where $P$ and $B$ stand for the pseudoscalar meson and baryon, respectively.

Note that the final states $K^{-}p$ are contributed with isospins both $I=0$ and $I=1$. We calculate the rescattering amplitudes in the isospin basis. For the isospin $I=0$, the four coupled channels $\bar{K}N$, $\pi\Sigma$, $\eta\Lambda$ and $K\Xi$ need to be considered. For the isospin $I=1$, there are five coupled channels $\bar{K}N$, $\pi\Sigma$, $\pi\Lambda$, $\eta\Sigma$ and $K\Xi$. Then we decompose the amplitudes of the physical states in Eq. (\ref{eq:amplitudes}) into the ones with the isospin states, shown as

\begin{equation}
	\begin{aligned} 
		&T_{\pi^{0}\Sigma^{0} \rightarrow K^{-}p}=-\frac{1}{\sqrt{6}}T_{\pi\Sigma \rightarrow \bar{K}N}^{I=0}, 
		\quad T_{\pi^{0}\Lambda \rightarrow K^{-}p}=-\frac{1}{\sqrt{2}}T_{\pi\Lambda \rightarrow \bar{K}N}^{I=1}, \\
		&T_{\eta\Sigma^{0} \rightarrow K^{-}p}=-\frac{1}{\sqrt{2}}T_{\eta\Sigma \rightarrow \bar{K}N}^{I=1}, 
		\quad T_{\eta\Lambda \rightarrow K^{-}p}=\frac{1}{\sqrt{2}}T_{\eta\Lambda \rightarrow \bar{K}N}^{I=0}, \\
		&T_{\pi^{+}\Sigma^{-} \rightarrow K^{-}p}=\frac{1}{2}T_{\pi\Sigma \rightarrow \bar{K}N}^{I=1}-\frac{1}{\sqrt{6}}T_{\pi\Sigma \rightarrow \bar{K}N}^{I=0}, \\
		&T_{K^{+}\Xi^{-} \rightarrow K^{-}p}=\frac{1}{2}T_{K\Xi \rightarrow \bar{K}N}^{I=1}-\frac{1}{2}T_{K\Xi \rightarrow \bar{K}N}^{I=0},
	\end{aligned}
	\label{eq:isospin}
\end{equation}

\noindent
where we have used the phase convention for the mesons $\left|\pi^{+}\rangle=-\right|1,1\rangle$, $\left|K^{-}\rangle=-\right| 1/2,-1/2\rangle$, and for the baryons $\left|\Sigma^{+}\rangle=-\right|1,1\rangle$, $\left|\Xi^{-}\rangle=-\right| 1/2,-1/2\rangle$ in terms of isospin states as used in Ref. \cite{Oset:1997it}.

In addition, the rescattering amplitude $T_{PB\rightarrow P^{'}B^{'}}$ can be obtained by solving the coupled channel Bethe–Salpeter equations of the on-shell form

\begin{equation}
	\begin{aligned} 
		T = [1-VG]^{-1}V, 
	\end{aligned}
	\label{eq:BSE}
\end{equation}

\noindent
where $G$ is a diagonal matrix composed of meson-baryon loop functions. The element of $G$ matrix with the dimensional regularization is given by \cite{Jido:2003cb}

\begin{equation}
	\begin{aligned}
		G_{PB}(M_{inv})=& \frac{2 m_{B}}{16 \pi^{2}}\left\{a_{PB}(\mu)+\ln \frac{m_{B}^{2}}{\mu^{2}}+\frac{m_{P}^{2}-m_{B}^{2}+M_{inv}^{2}}{2 M_{inv}^{2}} \ln\frac{m_{P}^{2}}{m_{B}^{2}}
		\right.\\&+\frac{q_{cm}(M_{inv})}{M_{inv}}\left[\ln \left(M_{inv}^{2}-\left(m_{B}^{2}-m_{P}^{2}\right)+2 q_{cm}(M_{inv}) M_{inv}\right)
		\right.\\&+\ln \left(M_{inv}^{2}+\left(m_{B}^{2}-m_{P}^{2}\right)+2 q_{cm}(M_{inv}) M_{inv}
		\right) \\&-\ln \left(-M_{inv}^{2}-\left(m_{B}^{2}-m_{P}^{2}\right)+2 q_{cm}(M_{inv}) M_{inv}
		\right) \\&\left.\left.-\ln \left(-M_{inv}^{2}+\left(m_{B}^{2}-m_{P}^{2}\right)+2 q_{cm}(M_{inv}) M_{inv}\right)\right]\right\},
	\end{aligned}
	\label{eq:GDR}
\end{equation}

\noindent
where $M_{inv}$ is the invariant mass of the meson–baryon system in the coupled channels, and $m_{P}$ $(m_{B})$ is the mass of the intermediate pseudoscalar meson (baryon). The $\mu$ is the scale of dimensional regularization, following Refs. \cite{Wang:2021lth, Oset:2001cn, Jido:2003cb}, which is taken as $0.63$ GeV, and $a_{PB}(\mu)$ is the subtraction constant, taken as

\begin{equation}
	\begin{aligned}
	a_{\bar{K} N}=-1.84, \quad a_{\pi \Sigma}=-2.00, \quad a_{\pi \Lambda}=-1.83, \\
	a_{\eta \Lambda}=-2.25, \quad a_{\eta \Sigma}=-2.38, \quad a_{K \Xi}=-2.67.
	\end{aligned}
	\label{eq:amu}
\end{equation}

\noindent
In order to study the properties of the intermediate resonances produced in the two-body interactions of the final states, these parameters are not regarded as free ones in our calculation, see Ref. \cite{Ahmed:2020qkv} for more discussions. Besides, $q_{cm}(M_{inv})$ is the three-momentum of the particle in the center-of-mass frame,

\begin{equation}
	\begin{aligned}
		q_{cm}(M_{inv})=\frac{\lambda^{1 / 2}\left(M_{inv}^{2}, m_{P}^{2}, m_{B}^{2}\right)}{2M_{inv}},
	\end{aligned}
	\label{eq:qcm}
\end{equation}

\noindent
with the usual Källen triangle function $\lambda(a, b, c)=a^{2}+b^{2}+c^{2}-2(a b+a c+b c)$.

Furthermore, the matrix $V$ denotes the $S$-wave interaction potentials for the coupled channels of $\bar{K}N$, of which the elements are taken from the lowest order chiral Lagrangian \cite{Pich:1995bw, Ecker:1994gg, Bernard:1995dp}. Finally the expressions for them are given by \cite{Oset:2001cn}

\begin{equation}
	V_{i j}(M_{inv})=-C_{i j} \frac{1}{4 f^{2}}\left(2 M_{inv}-m_{i}-m_{j}\right)\left(\frac{m_{i}+E_{i}}{2 m_{i}}\right)^{1 / 2}\left(\frac{m_{j}+E_{j}}{2 m_{j}}\right)^{1 / 2},
	\label{eq:Vij}
\end{equation}

\noindent
where the $m_{i}$, $m_{j}$ are the masses of the initial and final baryons, and $E_{i}$, $E_{j}$ the energies of the initial and final mesons. For the meson decay constant, we take $f=1.123f_{\pi}$, where the $f_{\pi}=0.093$ GeV is the pion decay constant. The coefficient matrix elements $C_{i j}$ are symmetric, $C_{ij}=C_{ji}$, given in Table \ref{tab:I0} for the $I=0$ sector, and in Table \ref{tab:I1} for the $I=1$ sector, which are taken from Ref. \cite{Oset:1997it}.

\begin{table}[htbp]
	\centering
	\caption{The coefficient matrix elements $C_{i j}$ in Eq. (\ref{eq:Vij}) for $I=0$ sector.}
	\resizebox{0.4\textwidth}{!}
	{\begin{tabular}{ccccc}
			\hline \hline 
			\quad$C_{i j}$&\quad$\bar{K}N$&\quad$\pi\Sigma$&\quad$\eta\Lambda$&\quad$K\Xi$ \\
			\hline 
			\quad$\bar{K}N$&\quad$3$&\quad$-\sqrt{\frac{3}{2}}$&\quad$\frac{3}{\sqrt{2}}$&\quad$0$ \\
			\quad$\pi\Sigma$&\quad&\quad$4$&\quad$0$&\quad$\sqrt{\frac{3}{2}}$ \\
			\quad$\eta\Lambda$&\quad&\quad&\quad$0$&\quad$-\frac{3}{\sqrt{2}}$ \\			
			\quad$K\Xi$&\quad&\quad&\quad&\quad$3$ \\
			\hline\hline  
	\end{tabular}}
	\label{tab:I0}
\end{table}

\begin{table}[htbp]
	\centering
	\caption{The coefficient matrix elements $C_{i j}$ in Eq. (\ref{eq:Vij}) for $I=1$ sector.}
	\resizebox{0.5\textwidth}{!}
	{\begin{tabular}{cccccc}
			\hline \hline 
			\quad$C_{i j}$&\quad$\bar{K}N$&\quad$\pi\Sigma$&\quad$\pi\Lambda$&\quad$\eta\Sigma$&\quad$K\Xi$ \\
			\hline 
			\quad$\bar{K}N$&\quad$1$&\quad$-1$&\quad$-\sqrt{\frac{3}{2}}$&\quad$-\sqrt{\frac{3}{2}}$&\quad$0$ \\
			\quad$\pi\Sigma$&\quad&\quad$2$&\quad$0$&\quad$0$&\quad$1$ \\
			\quad$\pi\Lambda$&\quad&\quad&\quad$0$&\quad$0$&\quad$-\sqrt{\frac{3}{2}}$ \\
			\quad$\eta\Sigma$&\quad&\quad&\quad&\quad$0$&\quad$-\sqrt{\frac{3}{2}}$ \\
			\quad$K\Xi$&\quad&\quad&\quad&\quad&\quad$1$ \\
			\hline\hline  
	\end{tabular}}
	\label{tab:I1}
\end{table}

\begin{figure}[htbp]
	\centering
	\includegraphics[width=0.5\linewidth,trim=150 540 180 130,clip]{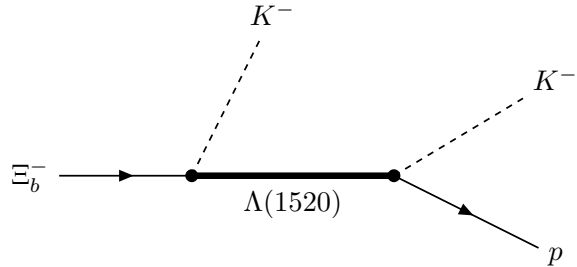} 
	\caption{The $\Xi_{b}^{-} \rightarrow pK^{-}K^{-}$ decay via the intermediate state $\Lambda(1520)$.}
	\label{fig:Lambda}
\end{figure} 

In addition, the $uds$ cluster in Fig. \ref{fig:Feynman1} can form intermediate particles directly and then decay into the final states $K^{-}p$. As implied in the  experimental results of the LHCb Collaboration \cite{LHCb:2021enr}, we consider the contributions of the intermediate state $\Lambda(1520)$ as shown in Fig. \ref{fig:Lambda}. The effective Lagrangians for the decay $\Xi_{b}^{-} \rightarrow \Lambda(1520)K^{-} \rightarrow pK^{-}K^{-}$ in Fig. \ref{fig:Lambda} are defined in general as follows \cite{Ahn:2019rdr}, of which the weak interaction vertex is given by

\begin{equation}
	\begin{aligned}
		\mathcal{L}_{K \Xi_{b} \Lambda^{*}}^{\text {weak }}=\frac{i}{m_{K}}\left(\bar{\Lambda}_{\mu}^{*} \right)\left(\partial^{\mu} K\right)\left(g_{K \Xi_{b} \Lambda^{*}}^{\mathrm{PV}}-g_{K \Xi_{b} \Lambda^{*}}^{\mathrm{PC}} \gamma_{5}\right) \Xi_{b}+\text {H.c.},
	\end{aligned}
	\label{eq:weakLagrangian}
\end{equation}

\noindent
where $g_{K \Xi_{b} \Lambda^{*}}^{\mathrm{PV}}$ and $g_{K \Xi_{b} \Lambda^{*}}^{\mathrm{PC}}$ are the parity violating (PV) and parity conserving (PC) couplings, respectively. To reduce theoretical uncertainties, we assume $g_{K \Xi_{b} \Lambda^{*}}^{\mathrm{PV}}=g_{K \Xi_{b} \Lambda^{*}}^{\mathrm{PC}}=g_{K \Xi_{b} \Lambda^{*}}$. The strong interaction vertex is written as

\begin{equation}
	\begin{aligned}
		\mathcal{L}_{K \Lambda^{*}p}^{\text {strong }}=-\frac{i g_{K\Lambda^{*}p}}{m_{K}}\left(\bar{\Lambda}_{\mu}^{*} \gamma_{5} \right)\left(\partial^{\mu} K\right)  p+\text {H.c.}.
	\end{aligned}
	\label{eq:strongLagrangian}
\end{equation}

\noindent
Using the above effective Lagrangians and the Breit-Wigner propagator \cite{Choi:1989yf, Sinha:2021mmx}, we can get the amplitude corresponding to Fig. \ref{fig:Lambda} as follows

\begin{equation}
	\begin{aligned}
		\mathcal{M}_{\Lambda^{*}}(M_{12})=-\frac{g_{K \Lambda^{*}p}}{m_{K}^{2}} \frac{\bar{u}_{p} \gamma_{5}\left[\Delta_{\mu \nu}\left(M_{12}\right) k_{2}^{\mu} k_{3}^{\nu}\right] \left(g_{K \Xi_{b} \Lambda^{*}}-g_{K \Xi_{b} \Lambda^{*}} \gamma_{5}\right) u_{\Xi_{b}^{-}}}{M_{12}^{2}-m_{\Lambda^{*}}^{2}+i \Gamma_{\Lambda^{*}} m_{\Lambda^{*}}},
	\end{aligned}
	\label{eq:amplitudeLambda1}
\end{equation}

\noindent
where $\Delta_{\mu \nu}(q)$ is the projection operator of spin $S=3/2$, given by

\begin{equation}
	\begin{aligned}
		\Delta_{\mu \nu}(q)=\left(\slashed{q}+m_{\Lambda^{*}}\right)\left[g_{\mu \nu}-\frac{1}{3} \gamma_{\mu} \gamma_{\nu}-\frac{1}{3 m_{\Lambda^{*}}}\left(\gamma_{\mu} q_{\nu}-\gamma_{\nu} q_{\mu}\right)-\frac{2}{3 m_{\Lambda^{*}}^{2}} q_{\mu} q_{\nu}\right].
	\end{aligned}
	\label{eq:Delta}
\end{equation}

\noindent
However, note that the $\Lambda(1520)$ state has the structure and can not be regarded as a point-like particle. Thus, we introduce the following form factor developed in Refs. \cite{Haberzettl:1998aqi,Davidson:2001rk} into the $\Lambda(1520)$ amplitude,

\begin{equation}
	\begin{aligned}
		F(M_{ij})=\frac{\Lambda^{4}}{\Lambda^{4}+\left(M_{ij}^{2}-m_{\Lambda(1520)}^{2}\right)^{2}},
	\end{aligned}
	\label{eq:FF}
\end{equation}

\noindent
where $\Lambda$ stands for a phenomenological cutoff parameter, which is taken as $1$ GeV \cite{Toki:2007ab}. In fact, this parameter has almost no effect on our fit, and it mainly influences the high-energy region. Taking into account the symmetry of identical particles $K^{-}K^{-}$ in the final states, we get the final amplitude for the $\Lambda(1520)$ contributions,

\begin{equation}
	\begin{aligned}
		\mathcal{M}_{\Lambda(1520)}(M_{12},M_{13})=-\frac{\mathcal{D}_{\Lambda(1520)}}{m_{K}^{2}} \frac{\bar{u}_{p} \gamma_{5} \left[\Delta_{\mu \nu}\left(M_{12}\right) k_{2}^{\mu} k_{3}^{\nu}\right] \left(1-\gamma_{5}\right) u_{\Xi_{b}^{-}}}{M_{12}^{2}-m_{\Lambda(1520)}^{2}+i \Gamma_{\Lambda(1520)} m_{\Lambda(1520)}}F(M_{12})+(2\leftrightarrow3),
	\end{aligned}
	\label{eq:amplitudeLambda2}
\end{equation}

\noindent
where $\mathcal D_{\Lambda(1520)}$ is a free parameter, which also can be determined by fitting the experimental data and have collected the couplings $g_{K \Xi_{b} \Lambda^{*}}$, $g_{K\Lambda^{*}p}$, and a global constant $C$ to match the events of the experimental data. Besides, the mass of $\Lambda(1520)$ is taken as $m_{\Lambda(1520)}=1.519$ GeV, and the width of the $\Lambda(1520)$ is taken as $\Gamma_{\Lambda(1520)}=0.016$ GeV, which are taken from the Particle Data Group (PDG) \cite{Zyla:2020}. Note that, the variables $M_{ij}$ are not completely independent, they fulfill the following constraint condition, which means that only two of them are independent,

\begin{equation}
	\begin{aligned}
		M_{12}^{2}+M_{13}^{2}+M_{23}^{2}=m_{\Xi_{b}^{-}}^{2}+m_{p}^{2}+m_{K^{-}}^{2}+m_{K^{-}}^{2}.
	\end{aligned}
	\label{eq:sij}
\end{equation}

Finally, the three-body double differential width distribution for the $\Xi_{b}^{-} \rightarrow pK^{-}K^{-}$ decay is given by \cite{Zyla:2020}

\begin{equation}
	\begin{aligned}
		\frac{d^{2} \Gamma}{d M_{12}d M_{13}}=\frac{1}{(2 \pi)^{3}} \frac{M_{12}M_{13}}{8 m_{\Xi_{b}^{-}}^{3}}\frac{1}{2} \left(\left|\mathcal{M}_{S-wave}\right|^{2}+\left|\mathcal{M}_{\Lambda(1520)}\right|^{2}\right),
	\end{aligned}
	\label{eq:dGamma}
\end{equation}

\noindent
where there is a factor $1/2$, since the two $K^{-}$ are identical. 
Since the $\Lambda(1520)$ contributes in $D$-wave, we take an incoherent sum for the contributions of the $S$-wave and the $\Lambda(1520)$ due to no interference between different partial waves under the orthogonality relation, where one should keep in mind that the scattering amplitudes of the coupled channels evaluated by Eq. \eqref{eq:BSE} are pure $S$-wave. This is different from the experimental modelling, where the nonzero unphysical interference would occur due to the symmetrization of the Dalitz plot as discussed in Ref. \cite{LHCb:2021enr}. Thus, these interference effects would lead to the source of systematic uncertainties. But, there is no such interference effect in our formalism. Furthermore, since the states $\Lambda (1405)$, $\Lambda (1670)$ and a new state with isospin $I=1$ (see the results later) are dynamically generated in the same sector of the coupled channel interactions, the interference effect has been contained in the scattering amplitudes, see Eq. \eqref{eq:amplitudes}, where more discussions can be found in Ref. \cite{Wang:2015pcn}.
In the present work, we aim at understanding the molecular nature of these $S$-wave low-lying resonances. Thus, we calculate the invariant mass spectrum $d\Gamma/dM_{12}$ by integrating the variable $M_{13}$ in Eq. (\ref{eq:dGamma}). In Sec. \ref{sec:Results}, we also evaluate the distribution $d\Gamma/dM_{23}$ through Eq. (\ref{eq:sij}).

\section{Results}
\label{sec:Results}

\begin{figure}[htbp]
	\centering
	\includegraphics[width=0.4\linewidth]{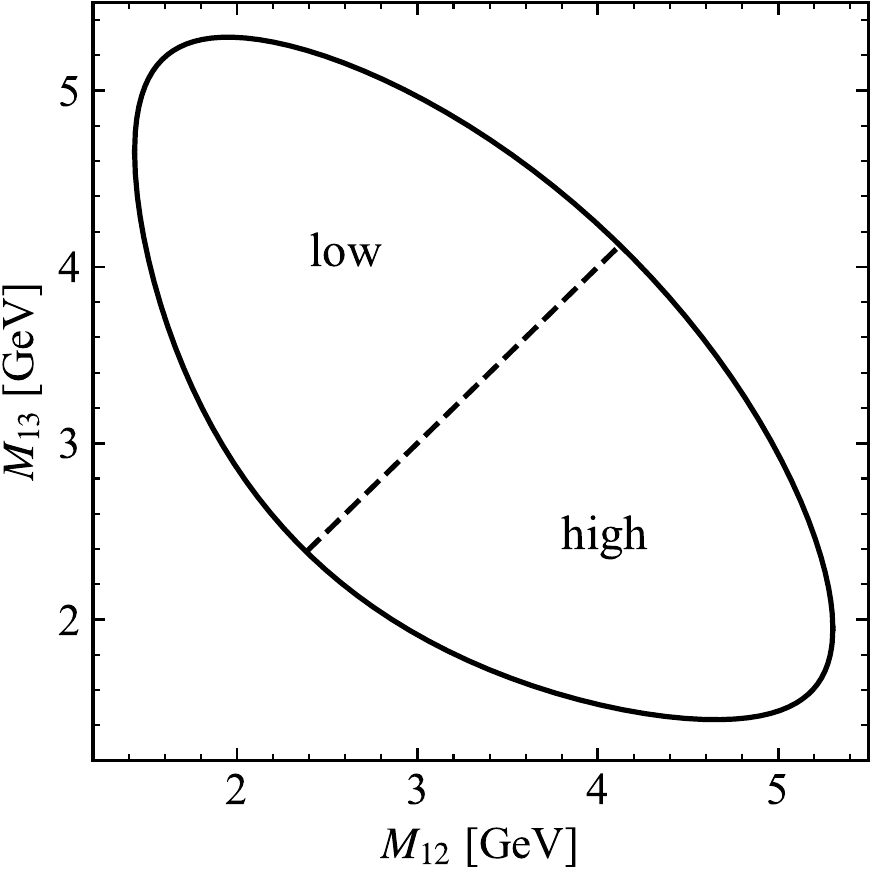} 
	\caption{Dalitz plot with the definitions of the regions $M_{12}^{low}$ and $M_{12}^{high}$.}
	\label{fig:Dalitzplot}
\end{figure}

As mentioned in the introduction, the LHCb Collaboration had measured the decay process $\Xi_{b}^{-} \rightarrow pK^{-}K^{-}$, where the $pK^{-}$ invariant mass spectrum was given. 
Note that the two identical $K^{-}$ mesons lead to $M_{12}$ and $M_{13}$ having a symmetry under interchanging the variables, thus the experimental results are described by $M_{pK^{-}}^{low}$ and $M_{pK^{-}}^{high}$ variables, which are the lower and higher values among $M_{12}$ and $M_{13}$ \cite{LHCb:2019tdw} due to their  different energies. Therefore, when fitting the invariant mass spectrum of $pK^{-}$, the limits of the integral in Eq. (\ref{eq:dGamma}) are described by Fig. \ref{fig:Dalitzplot}, which are similar to what had been done in Ref. \cite{Roca:2020lyi}. In fact, due to two identical kaons, the regions $M_{12}^{low}$ and $M_{12}^{high}$ as shown in Fig. \ref{fig:Dalitzplot} could be also for the variable of $M_{13}$. Thus, when one folds the symmetry parts of Fig. \ref{fig:Dalitzplot} along $M_{12}=M_{13}$, all the data can be described by the folded Dalitz plot, which is really done in the experiments \cite{LHCb:2021enr} with $m_{low}$ and $m_{high}$.

\begin{figure}[htbp]
	\begin{subfigure}{0.47\textwidth}
		\centering
		\includegraphics[width=1\linewidth]{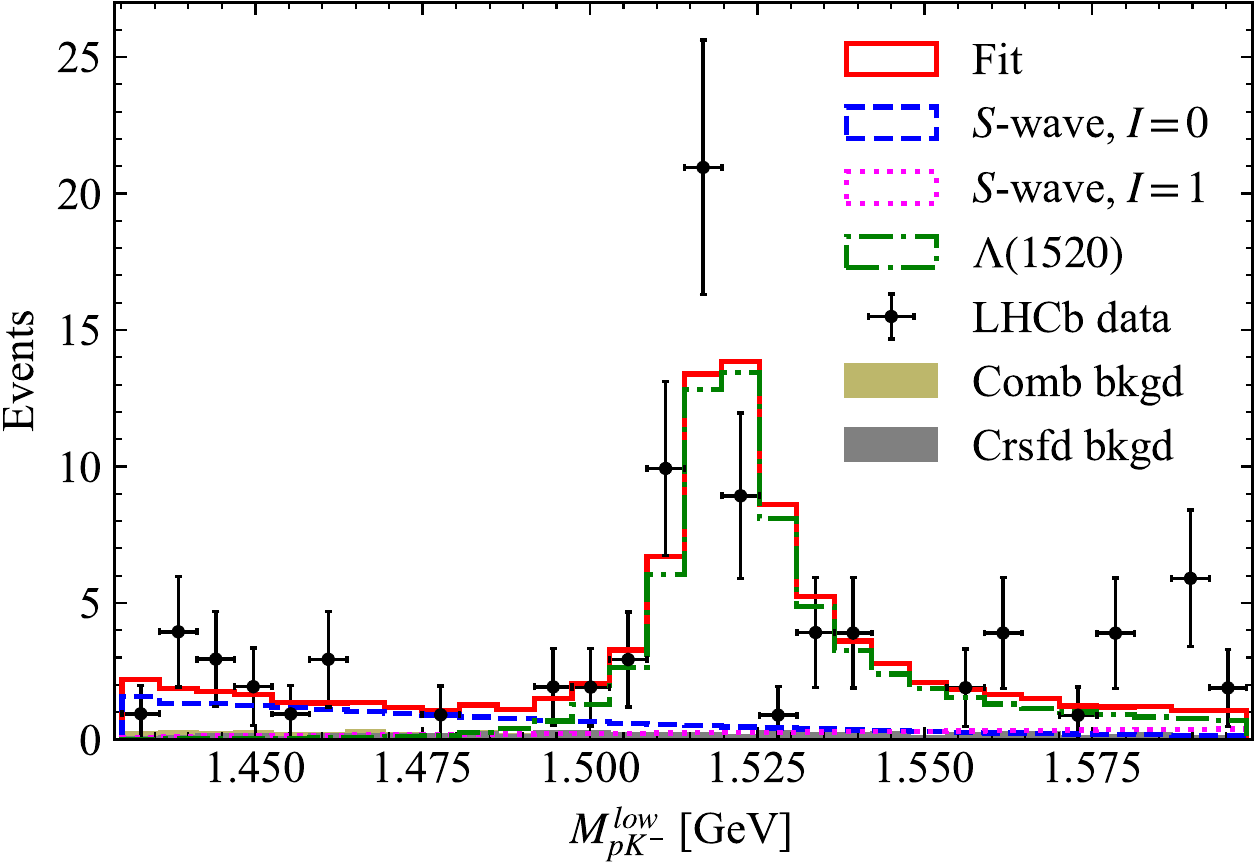} 
		\caption{The invariant mass distributions of lower $pK^{-}$ in $[1.43, 1.60]$ GeV.}
		\label{fig:MinvpKlow7a}
	\end{subfigure}
	\quad
	\quad
	\begin{subfigure}{0.47\textwidth}  
		\centering 
		\includegraphics[width=1\linewidth]{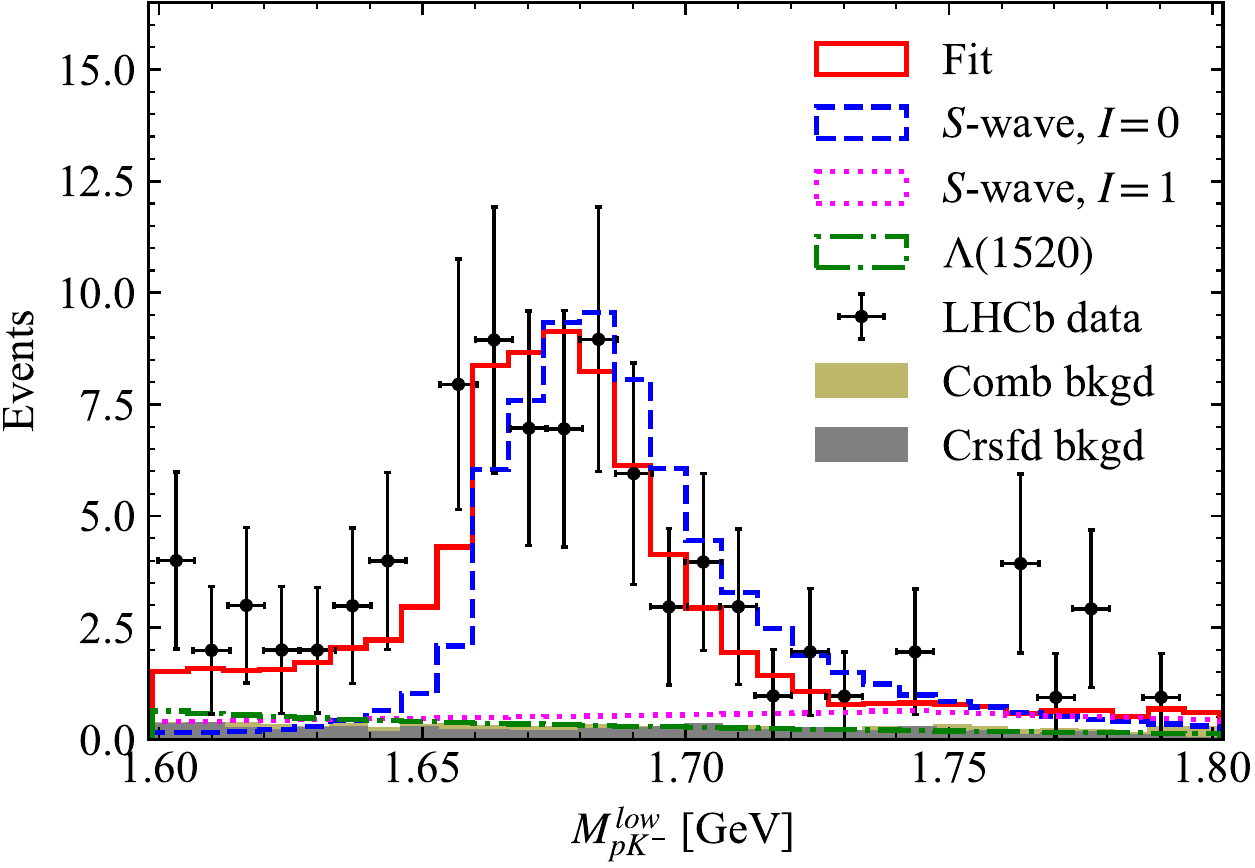} 
		\caption{The invariant mass distributions of lower $pK^{-}$ in $[1.60, 1.80]$ GeV.}
		\label{fig:MinvpKlow8a}  
	\end{subfigure}	
	\caption{The invariant mass distributions of lower $pK^{-}$ in the $\Xi_{b}^{-} \rightarrow pK^{-}K^{-}$ decay. The solid (red) line is the total contributions of the $S$-wave with isospin $I=0$ and $I=1$, the $\Lambda(1520)$ and the background. The dashed (blue) and dotted (magenta) lines are the contributions from the $S$-wave with isospins $I=0$ and $I=1$, respectively. The dash-dot (green) line is the $\Lambda(1520)$ contributions. The dots (black) are the LHCb experimental data, the dark khaki and grey histograms are the combinatorial (Comb) and cross feed (Crsfd) backgrounds, respectively, which all are taken from Ref. \cite{LHCb:2021enr}.}
	\label{fig:MinvpKlow7a8a}
\end{figure}

In our calculation, only two free parameters need to be determined by fitting the experimental data, i.e., $\mathcal D$ and $\mathcal{D}_{\Lambda(1520)}$, which represent the strength of the $S$-wave FSI and the $\Lambda(1520)$ in $D$-wave, respectively. They are uncorrelated and do not affect the line shape of theirs invariant mass spectra. 
First, we make a combined fit of the LHCb experimental data as shown Fig. \ref{fig:MinvpKlow7a8a}, and the fitted parameters and $\chi^{2}/dof.$ are given in Table \ref{tab:Parameters}. We can see that our results are in agreement with the experimental data. Note that we only use one set of (two) parameters, see Table \ref{tab:Parameters}, and obtain good description for two sets of experimental data of Figs. \ref{fig:MinvpKlow7a} and \ref{fig:MinvpKlow8a}. 
In Fig. \ref{fig:MinvpKlow7a}, above the $pK^{-}$ threshold, the contributions from the resonance $\Lambda(1405)$ are generated by the coupled channel interactions of the $S$-wave with isospin $I=0$ using the ChUA, which are shown by the dashed (blue) line. In the middle-energy region, as shown by the dash-dot (green) line, the structure of the $\Lambda(1520)$ state is clear. As shown in Fig. \ref{fig:MinvpKlow8a} for the lower $pK^{-}$ invariant mass distributions from $1.6$ to $1.8$ GeV, the contributions from the $\Lambda(1670)$ state is particularly visible, which is also dynamically generated in the $S$-wave FSI, recalling that there is no contribution from the tree level diagram. As analyzed in Ref. \cite{Wang:2021lth}, the $\Lambda(1405)$ corresponds to two states, one with higher-mass is a pure $\bar KN$ molecule, the other one is a compositeness of main components of $\pi\Sigma$ and small part of $\bar KN$, while the $\Lambda(1670)$ is a bound state of $K\Xi$. In Fig. \ref{fig:MinvpKlow8a}, the peak structures in the total (solid, red) line and the one (dashed, blue) in the $S$-wave FSI with isospin $I=0$ have obvious horizontal dislocation, while the $\Lambda(1520)$ has almost no contribution. This differences between them indicate that there will be another resonance contributed in the $S$-wave FSI with isospin $I=1$, see the dotted (magenta) line and the following analysis.

\begin{table}[htbp]
	\centering
	\caption{Values of the parameters from the fit.}
	\resizebox{0.8\textwidth}{!}
	{\begin{tabular}{cccc}
			\hline \hline 
			Parameters &\quad $\mathcal D$&\quad $\mathcal D_{\Lambda(1520)}$&\quad $\chi^{2}/dof.$   \\
			\hline 
			Fit results &\quad $810.08\pm22.79$&\quad $1.70\pm0.04$&\quad $86.38/(46-2)=1.96$ \\
		
			\hline\hline  
	\end{tabular}}
	\label{tab:Parameters}
\end{table}

\begin{figure}[htbp]
	\begin{subfigure}{0.47\textwidth}
		\centering
		\includegraphics[width=1\linewidth]{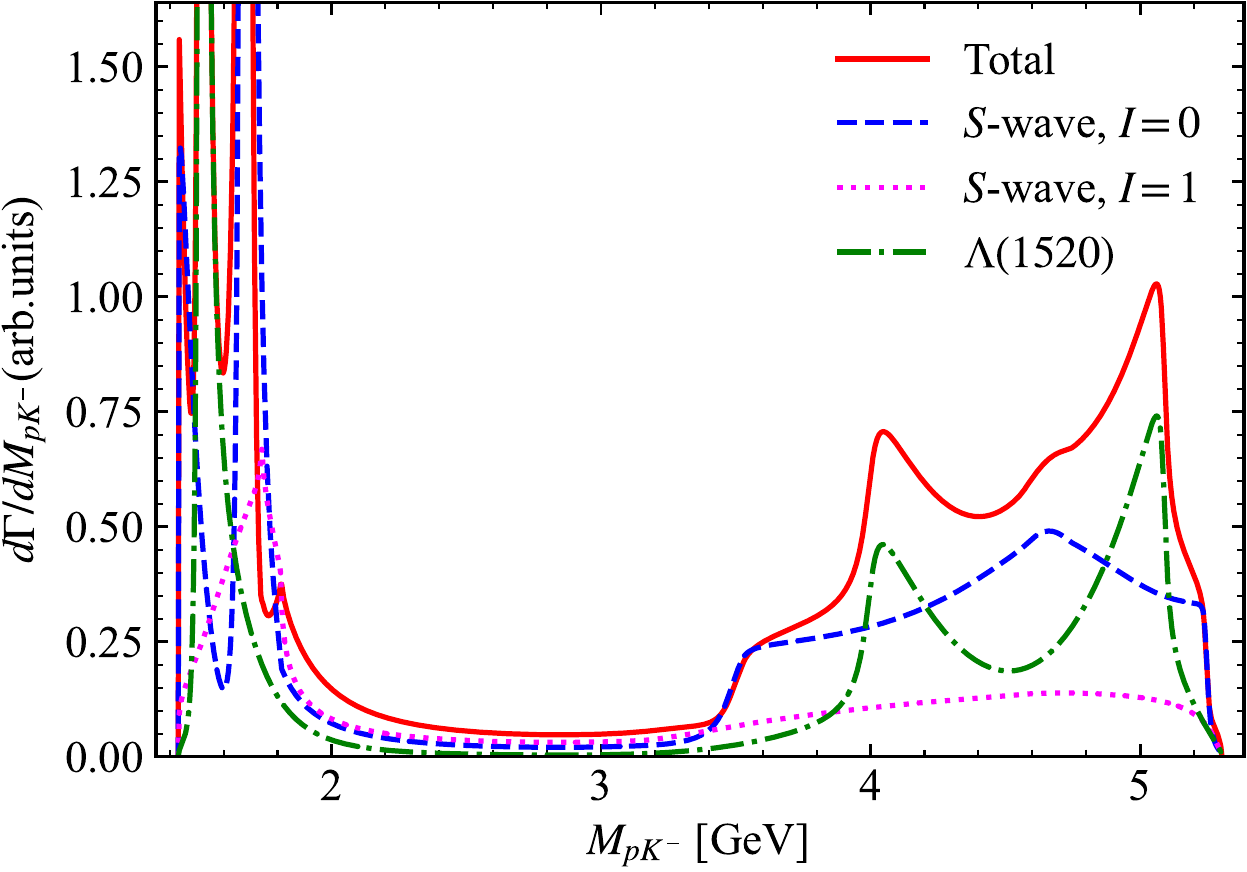} 
		\caption{The invariant mass distributions of $pK^{-}$.}
		\label{fig:MinvpK}
	\end{subfigure}
	\quad
	\quad
	\begin{subfigure}{0.47\textwidth}  
		\centering 
		\includegraphics[width=1\linewidth]{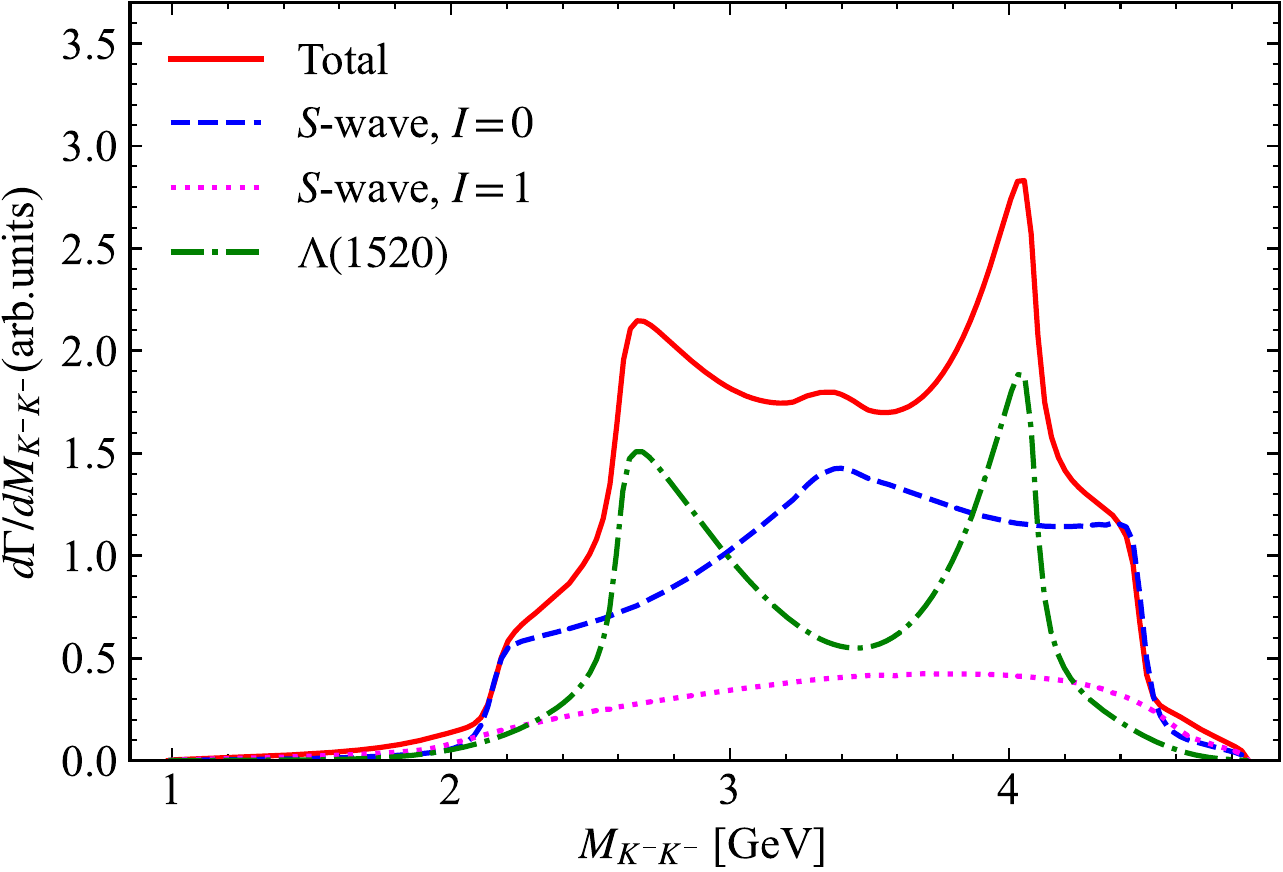} 
		\caption{The invariant mass distributions of $K^{-}K^{-}$.}
		\label{fig:MinvKK}  
	\end{subfigure}	
	\begin{subfigure}{0.47\textwidth}  
		\centering 
		\includegraphics[width=1\linewidth]{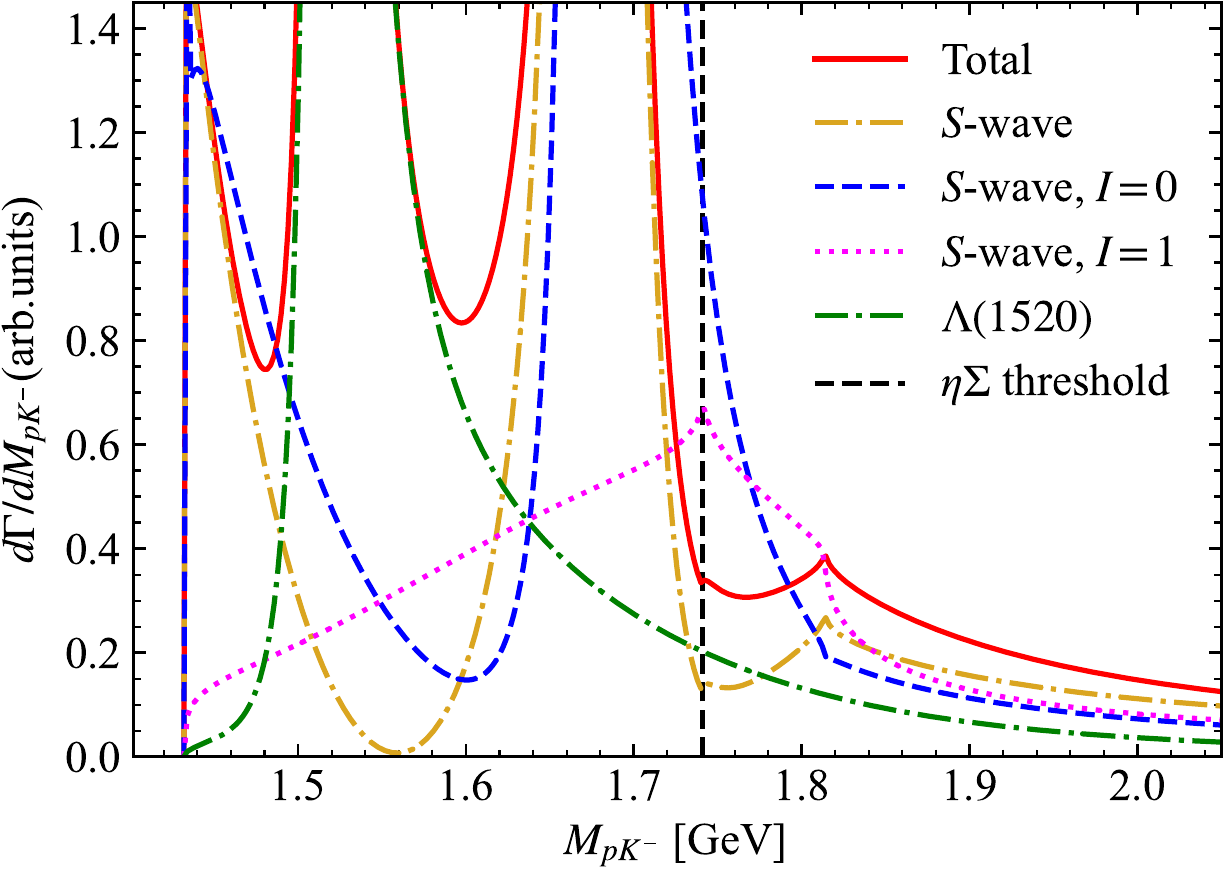} 
		\caption{The invariant mass distributions of $pK^{-}$ in the low-energy region.}
		\label{fig:MinvpKpart}  
	\end{subfigure}
	\caption{The invariant mass distributions of $pK^{-}$ and $K^{-}K^{-}$ for the $\Xi_{b}^{-} \rightarrow pK^{-}K^{-}$ decay. The results for different lines are the same as Fig. \ref{fig:MinvpKlow7a8a}. Furthermore, the dash-dot (goldenrod) line is the total contributions of the $S$-wave with isospins $I=0$ and $I=1$, the dashed (black) vertical line is the threshold of $\eta\Sigma$ channel.}
	\label{fig:Minv}
\end{figure}

Next, we plot the $pK^{-}$ and $K^{-}K^{-}$ invariant mass distributions in the full energy regions for the decay of $\Xi_{b}^{-} \rightarrow pK^{-}K^{-}$, see the results of Fig. \ref{fig:Minv}. 
Thus, the horizontal axis of $M_{pK^-}$ in Fig. \ref{fig:MinvpK} is in fact the full energy range of $M_{12}$, $M_{12}^{low}$ + $M_{12}^{high}$, whereas, the one in Fig. \ref{fig:MinvpKpart} is for the region of $M_{12}^{low} < 2.05$ GeV. The results of Fig. \ref{fig:MinvpK} are consistent with the line shapes as shown in Figs. 7-11 of Ref. \cite{LHCb:2021enr}, where we do not compare these experimental results with events one parts by one parts due to the higher energy region out of our concern as discussed in the introductions. 
In Fig. \ref{fig:MinvpK}, the peak structures near the $pK^{-}$ threshold are contributed by the amplitudes of the $S$-wave FSI and the $D$-wave $\Lambda(1520)$, and the structures in the high-energy region are caused by the reflections of the resonance structures appeared in the low-energy region. Especially, it is not surprising that there are two-peak structures for the contributions from the $D$-wave $\Lambda(1520)$, where a Breit-Wigner type amplitude is taken for its contribution of the $D$-wave, see Eq. \eqref{eq:amplitudeLambda2}, and which were also shown in the fitting results of Refs. \cite{LHCb:2021enr,Wang:2021ews,Wang:2021kka}. Indeed, the Breit-Wigner type amplitude will also show up some peak structures when it is projected to the higher energy region or the other energy variable, see more discussions in Ref. \cite{Wang:2020dmv}. 
For the invariant mass spectrum of $K^{-}K^{-}$ in Fig. \ref{fig:MinvKK}, the main contributions also come from the reflections of the amplitudes of the $\Lambda(1520)$ and the FSI with isospins $I=0,\ 1$, of which the line shape is similar to the part in the high-energy region of Fig. \ref{fig:MinvpK}. 
Figure \ref{fig:MinvpKpart} shows more detailed structures in the low-energy region of Fig. \ref{fig:MinvpK}. In Fig. \ref{fig:MinvpKpart}, except for the obvious structures of the states $\Lambda(1405)$, $\Lambda(1520)$ and $\Lambda(1670)$, there is another resonance appearing in the FSI with isospin $I=1$. The peak of this state is close to the threshold of $\eta\Sigma$ channel and has a typical cusp effect, which spans from the $pK^{-}$ threshold to about $2$ GeV, and leads to a small horizontal dislocation in the line shape of the total invariant mass spectrum and the one of $S$-wave FSI with isospin $I=0$, see the solid (red) and dashed (blue) lines, respectively, as shown in Figs. \ref{fig:MinvpKlow8a} and \ref{fig:MinvpKpart}. 
As done in Ref. \cite{Wang:2021lth} to extrapolate the scattering amplitudes to the general second Riemann sheet, we found a pole $\sqrt{s_{p}}=(1579.52+264.40i)$ MeV in the isospin $I=1$ sector, which has total angular momentum $J=1/2$ and is in agreement with the one $\sqrt{s_{p}}=(1579+264i)$ MeV found in Refs. \cite{Oset:2001cn,Ramos:2003mu,Dong:2016auh}. 
Unfortunately, this resonance was not found in the experiments \cite{LHCb:2021enr}. 
In fact, this new state with $I=1$ is difficult to be detected in the present $\Xi_{b}^{-}$ decay process, since its signal is destroyed totally by the interference effects from the states $\Lambda(1405)$ and $\Lambda(1670)$, as shown in the total results of Fig. \ref{fig:MinvpKpart}. 
Since this state is below the $\eta\Sigma$ threshold, which couples strongly to the $K\Xi$ channel \cite{Oset:2001cn,Ramos:2003mu,Dong:2016auh} and can decay into the $\bar{K} N$, $\pi \Sigma$ and $\pi \Lambda$ channels in the isospin $I=1$ sector. Therefore, some decay processes of the heavy hadrons, such as $\Lambda_b$ and $\Xi_b$ baryons, $B$ and $B_s$ mesons, which contain these final states of $\bar{K} N$, $\pi \Sigma$ and $\pi \Lambda$, may be detected for this new state. 
Thus, more accurate experimental results are needed to find it in the future.

Furthermore, we calculate the branching fractions of the corresponding decay channel. In our model, we do not know the weak interaction vertex factor $V_{P}$ in Eq. (\ref{eq:H3}) and the couplings in Eq. (\ref{eq:amplitudeLambda1}). Therefore we use the experimentally measured branching fraction of the decay channel via $\Lambda(1520)$ as the known input to calculate the branching fractions of the decay channels via other resonances. In the evaluations, except for the vertex factor and the couplings, there is also a global constant $C$, which can be eliminated by calculating the ratio as below. By integrating the $pK^{-}$ invariant mass distributions of the dashed (blue) line for the $\Lambda(1405)$ and the dash-dot (green) line for the $\Lambda(1520)$ in Fig. \ref{fig:MinvpKpart}, we obtain the ratio,

\begin{equation} 
	\begin{aligned}
		\frac{\mathcal{B}(\Xi_{b}^{-} \rightarrow \Lambda(1405)K^{-}, \Lambda(1405) \rightarrow pK^{-})}{\mathcal{B}(\Xi_{b}^{-} \rightarrow \Lambda(1520)K^{-}, \Lambda(1520) \rightarrow pK^{-})} = 0.24^{+0.04}_{-0.02},
	\end{aligned}
	\label{fig:Ratios1}
\end{equation} 

\noindent
where the integrate limits for $\Xi_{b}^{-} \rightarrow \Lambda(1405)K^{-}$ decay are taken from the $pK^{-}$ threshold up to $1.6$ GeV, and the ones for $\Xi_{b}^{-} \rightarrow \Lambda(1520)K^{-}$ decay from the $pK^{-}$ threshold up to $1.8$ GeV. The uncertainties come from the changes of upper limits $1.6\pm0.05$ GeV and $1.8\pm0.05$ GeV in the upper limits. Analogously, we get the following fractions

\begin{equation} 
	\begin{aligned}
		\frac{\mathcal{B}(\Xi_{b}^{-} \rightarrow \Lambda(1670)K^{-}, \Lambda(1670) \rightarrow pK^{-})}{\mathcal{B}(\Xi_{b}^{-} \rightarrow \Lambda(1520)K^{-}, \Lambda(1520) \rightarrow pK^{-})} = 1.12 ^{+0.03}_{-0.04},
	\end{aligned}
	\label{fig:Ratios2}
\end{equation} 

\begin{equation} 
	\begin{aligned}
		\frac{\mathcal{B}(\Xi_{b}^{-} \rightarrow RK^{-}, R \rightarrow pK^{-})}{\mathcal{B}(\Xi_{b}^{-} \rightarrow \Lambda(1520)K^{-}, \Lambda(1520) \rightarrow pK^{-})} = 0.40 ^{+0.01}_{-0.01},
	\end{aligned}
	\label{fig:Ratios3}
\end{equation} 

\noindent
where $R$ represents the state from FSI with isospin $I=1$ as discussed above. The integrate limits for the $\Xi_{b}^{-} \rightarrow \Lambda(1670)K^{-}$ decay taken from $1.6$ up to $1.9$ GeV, where the uncertainties come from the changes of lower limits $1.6\pm0.05$ GeV. The ones for the $\Xi_{b}^{-} \rightarrow RK^{-}$ decay from the $pK^{-}$ threshold up to $2$ GeV, where the uncertainties come from the changes of upper limits $2\pm0.05$ GeV. The decay branching fraction measured by the LHCb Collaboration experiment is $\mathcal{B}(\Xi_{b}^{-} \rightarrow \Lambda(1520)K^{-}, \Lambda(1520) \rightarrow pK^{-})=(7.6\pm0.9\pm0.8\pm3.0)\times10^{-7}$ \cite{LHCb:2021enr}, and then combining the above values in Eqs. (\ref{fig:Ratios1})-(\ref{fig:Ratios3}), we obtain the other three branching fractions

\begin{equation} 
	\begin{aligned}
		\mathcal{B}(\Xi_{b}^{-} \rightarrow \Lambda(1405)K^{-}, \Lambda(1405) \rightarrow pK^{-})=(1.80\pm0.76^{+0.30}_{-0.18})\times10^{-7}, \\
		\mathcal{B}(\Xi_{b}^{-} \rightarrow \Lambda(1670)K^{-}, \Lambda(1670) \rightarrow pK^{-})=(8.51\pm3.62^{+0.20}_{-0.33})\times10^{-7}, \\
		\mathcal{B}(\Xi_{b}^{-} \rightarrow RK^{-}, R \rightarrow pK^{-})=(3.08\pm1.31^{+0.08}_{-0.10})\times10^{-7},
	\end{aligned}
	\label{fig:Theory}
\end{equation}

\noindent
where the first uncertainties are estimated from the experimental errors of  $\mathcal{B}(\Xi_{b}^{-} \rightarrow \Lambda(1520)K^{-}, \Lambda(1520) \rightarrow pK^{-})$, and the second ones are estimated from the integrations in Eqs. (\ref{fig:Ratios1})-(\ref{fig:Ratios3}). The following results are measured by the LHCb Collaboration \cite{LHCb:2021enr},

\begin{equation} 
	\begin{aligned}
		\mathcal{B}(\Xi_{b}^{-} \rightarrow \Lambda(1405)K^{-}, \Lambda(1405) \rightarrow pK^{-})=(1.9\pm0.6\pm0.7\pm0.7)\times10^{-7}, \\
		\mathcal{B}(\Xi_{b}^{-} \rightarrow \Lambda(1670)K^{-}, \Lambda(1670) \rightarrow pK^{-})=(4.5\pm0.7\pm1.3\pm1.8)\times10^{-7}.
	\end{aligned}
	\label{fig:LHCb}
\end{equation}

\noindent
For the $(\Xi_{b}^{-} \rightarrow \Lambda(1405)K^{-}, \Lambda(1405) \rightarrow pK^{-})$ decay, the theoretical value is consistent with the experiment within the uncertainties. We can see that even though the central values of our branching fraction of $(\Xi_{b}^{-} \rightarrow \Lambda(1670)K^{-}, \Lambda(1670) \rightarrow pK^{-})$ decay is about two times larger than the one of the LHCb Collaboration measured, our result is consistent with the measurement within the uncertainties. For the predicted branching ratio corresponding to the resonance generated by the FSI with isospin $I=1$, further experimental measurements are hopefully performed.
 
\section{Conclusions}
\label{sec:Conclusions}

The three-body decay of $\Xi_{b}^{-} \rightarrow pK^{-}K^{-}$ is studied by taking into account the final state interactions based on the chiral unitary approach. The dominant Feynman diagram contributions from the $W$-external emission mechanism are considered in the weak decay process. Our analysis shows that the final states $pK^{-}$ can not be directly produced in the $S$-wave at the tree level, and the rescattering effect of the final states is mandatory. We also take into account the contributions from the state $\Lambda(1520)$ using the corresponding effective Lagrangian. Our fitting results for the invariant mass distributions of lower $pK^{-}$ are consistent with the experimental data. Then we present the detailed $pK^{-}$ invariant mass spectra, where the resonances $\Lambda(1405)$ and $\Lambda(1670)$ are dynamically reproduced in the $S$-wave final state interactions with the isospin $I=0$, which indicates the molecular nature of these two states.  In addition, we find the contributions in the invariant mass distributions from a structure with isospin $I=1$, of which the pole is located at $\sqrt{s_{p}}=(1579.52+264.40i)$ MeV. However, this state has not yet been observed experimentally. As discussed in Refs. \cite{Oset:2001cn,Ramos:2003mu,Dong:2016auh}, this resonance is strongly coupled to the $K\Xi$ channel, too. Furthermore, we calculate the branching ratios of the corresponding decay channels. The result of the branching fraction $\mathcal{B}(\Xi_{b}^{-} \rightarrow \Lambda(1405)K^{-}, \Lambda(1405) \rightarrow pK^{-})$ is consistent with the measurements of the LHCb Collaboration within the uncertainties, while the one of $\mathcal{B}(\Xi_{b}^{-} \rightarrow \Lambda(1670)K^{-}, \Lambda(1670) \rightarrow pK^{-})$ is a little bigger than theirs. We hope that the future experiments could search for the predicted resonance around $1580$ MeV with isospon $I=1$ and make further measurements for its corresponding branching fraction.

\section*{Acknowledgements}

We would like to thank En Wang and Pei-Rong Li for valuable discussions. This work is supported by the China National Funds for Distinguished Young Scientists under Grant No. 11825503, National Key Research and Development Program of China under Contract No. 2020YFA0406400, the 111 Project under Grant No. B20063, the National Natural Science Foundation of China under Grant No. 12047501, and by the Fundamental Research Funds for the Central
Universities.

 \addcontentsline{toc}{section}{References}
\end{document}